\documentclass[epj]{svjour} %{svjour2}

\usepackage{color}
\usepackage{graphics}
\usepackage{amsmath}
\usepackage{amssymb}
\begin{document}
\title{Cell survival probability in a spread-out Bragg peak for novel treatment planning}
%\title{How to make the cell survival probability uniform along a spread-out Bragg peak}
\author{Eugene Surdutovich\inst{1,2}
    \and Andrey V. Solov'yov\inst{2,3}
}                     % Do not remove
\offprints{Eugene Surdutovich}          % Insert a name or remove this line
\institute{Department of Physics, Oakland University, Rochester, Michigan 48309, USA
     \and
MBN Research Center,
% Altenhoferallee 3
60438 Frankfurt am Main, Germany
\and On leave from A.F. Ioffe Physical Technical Institute, % Politechnicheskaya 26,
194021 St. Petersburg, Russian Federation}
\authorrunning{E. Surdutovich and A.V. Solov'yov}
\titlerunning{Cell survival along SOBP}

\date{Received: date / Revised version: date}
% The correct dates will be entered by Springer
%
\abstract{The problem of variable cell survival probability along the spread-out Bragg peak is one of the long standing problems in planning and optimisation of ion-beam therapy. This problem is considered using the multiscale approach to the physics of ion-beam therapy. The physical reasons for this problem are analysed and understood on a quantitative level. A recipe of solution to this problem is suggested using this approach. This recipe can be used in the design of a novel treatment planning and optimisation based on fundamental science.
} %end of abstract
\maketitle

The ion-beam cancer therapy has been developed in the 1990s as a hopeful improvement of a conventional radiation therapy with x-rays. The possibility of employment of a physical behaviour of ionization cross section of ions at decreasing energies giving rise to the Bragg peak in the depth-dose curve for better focusing the locus of radiation damage and sparing healthy tissues is the main attraction of this endeavour~\cite{Kraft07,Pedroni05,Hiroshiko,SchardtRMP10,Durante10,FokasKraft09,Hall,IBCTbook}.
Proton-beam therapy is the most proliferated type of ion-beam therapy, and this paper is more linked with protons, however, the results can be easily extended to include heavier ions.

\begin{figure}
\begin{centering}
\resizebox{0.99\columnwidth}{!}
{\includegraphics{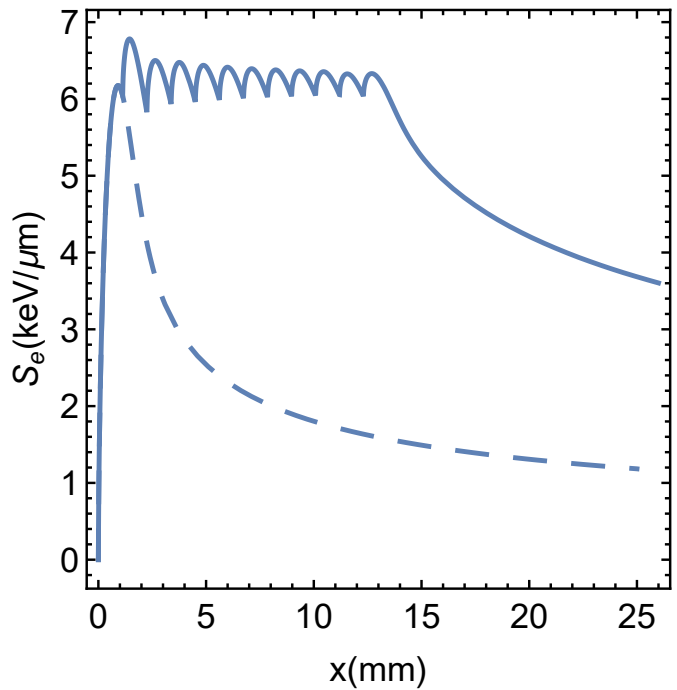}}
\caption{\label{fig.bragg} The dependence of LET for protons as a function of distance from the distal side of depth-dose curve. The dashed line shown a pristine Bragg peak calculated for protons with an initial energy of 200~MeV. The solid line shows a SOBP obtained as a superposition of proton beams of initial energies from 194.5 to 200~MeV with fluences chosen so that the physical dose profile along the SOBP is uniform.}
\end{centering}
\end{figure}

When proton-beam therapy is used, a beam comprised of protons of a given energy enters a patient's tissue. Typical initial energies are in the range of 70-230~MeV. Having entered the tissue, protons gradually slow down until their energy is above $\sim1$~MeV; at a high speed the probability of interaction with the medium is small and the depth-dose curve exhibits a plateau. When protons energy drops below 1~MeV (1~MeV/u for heavier ions), this probability (related to the cross section of inelastic collisions with molecules of tissue) increases rapidly. As a result projectiles lose most of their remaining energy within a short length segment referred to as the pristine Bragg peak. The position of the pristine Bragg peak for a given medium (such as tissue) and type of ions solely depends on their initial energy. Then, tuning the initial energy so that the position of the Bragg peak overlaps with a tumour location seems to be the next logical step. However, typical tumour sizes exceed the 1-mm size of the Bragg peak by at least an order of magnitude. Then one have to scan the tumour with beams of different energies; this creates another phenomenon, the spread-out Bragg peak (SOBP). Images of pristine and spread out Bragg peaks are shown in Fig.~\ref{fig.bragg}.

Radiation oncologists or medical physicists involved in planning and optimization of irradiation of patients base their protocols on a relation of a necessary dose to be delivered to a given voxel to the desired probability of cell survival in that volume. The dependence of this probability on dose is determined by the so-called survival curves. %The latter are a subject of a scientific debate.
In general, the survival curves depend on the type of projectiles, type of cells, phase of cells in their cycle, degree of oxygenation, and other possible conditions~\cite{Hall}. In the case of x-ray radiation, the only radiation parameter that describes the irradiation of a voxel is the dose, even though the mechanisms of radiation damage are rather complex. The target dose for treatment planning can be determined given the experimentally known survival curve for given cells and conditions~\cite{Hall,Alpen}.

In the case of irradiation with ions, there are at least two major issues. They originate from the fact that the dose, which physically corresponds to energy, loses its uni\-queness in predicting biodamage. It becomes inevitable to address the fact that the damage is not done by {\em energy}; instead, it is due to physical and chemical processes caused by agents such as electrons ejected by ions, free radicals (also formed as a result of the action of ions), other reactive species, and possibly other physical effects. { This is reflected in the introduction of such macroscopic parameters as relative biological effectiveness (RBE) and oxygen enhancement ratio(OER).} The abundance and spatial distribution of the above reactive species depend on dose and linear energy transfer (LET), the medium, etc., but this dependence is not trivial; the understanding of radiation biodamage on a fundamental level is necessary in order to establish this dependence and bring therapy optimisation and planning to a higher scientific level.

The radiation quality, a term related to the value of LET, of the delivered dose depends on the position along the SOBP. The cell survival depends on both the dose and the LET. The distal end is irradiated with protons at their pristine Bragg peak while the proximal end is irradiated with a superposition of protons with different values of LET; only a few of them are at the Bragg peak energies. This introduces a difference between the quality of different parts of the SOBP leading to different values of both relative biological effectiveness (RBE) and oxygen enhancement ratio (OER). The existence of this problem has been recognised a long time ago and there have many studies devoted to this. A thorough experimental analysis has been done and discussed, e.g., in ref.~\cite{Paganetti}.

The currently accepted model for relating the dose with cell survival probability for ions is the local effect model (LEM)~\cite{LEM96,SchardtRMP10,Kraemer,Elsaesser}), which associates the dose delivered to a voxel calculated probabilistically with the probability of cell survival taken from x-ray survival curves. This model only accounts for the effects of dose deposited in cell nuclei without analyzing physical, chemical or biological effects. Although this model claims predictive results on the basis of well known x-ray survival curves, there are serious doubts that these survival curves are relevant in the case of ions~\cite{Beuve,MSAColl}. More recent versions of LEM model are available\cite{LEMcluster,lem2012}, but they account for quality only empirically, again tailoring the radiation quality effects with x-ray curves.
On top of this, there is no solid scientific assessment of radiation damage inside the SOBP, since the cell survival depends on LET and, hence, on the position in the SOBP~\cite{Paganetti}. In this paper we will argue that the lack of understanding of this dependence on LET is the main reason for the problem of varying cell survival along the SOBP has not yet been solved~\cite{Paganetti16}. This topic is discussed at numerous conferences, such as Radiation Research Society (RRS), Application of Accelerators in Research and Industry (CAARI), Particle Therapy Co-operative Group (PTCOG), to name a few.

The problem of varying cell survival along the SOBP is vexing for proton-beam therapy and it is important to solve it. In this paper, we present a recipe of solution to this problem using the developed multiscale approach (MSA) to the physics of radiation damage with ions~\cite{pre,MSAColl}. The MSA relies directly on the physical, chemical, and biological effects that underlie the observed biological damage. In this approach, a target cell nucleus is irradiated with ions having different LET. Then the average number of lethal lesions per cell is calculated taking into account the LET, cell properties, and external conditions such as degree of oxygenation. This way, the dose and the radiation quality are directly associated with a predicted radiation damage. The predictability of cell survival by the MSA has been successfully tested on a variety of cell lines with different values of LET and oxygenation conditions~\cite{CellSurSR16}. It is tempting to use the technique vindicated by that work to develop an algorithm of obtaining a flat profile for the cell survival along the SOBP.

%We argue that this method is very well suited for development of the algorithm of determining the desired pattern of irradiation with protons in order to achieve a uniform distribution of a given survival probability over the target, since the position within the SOBP can be included as an independent variable determining the average number of lethal lesions. Besides, the oxygen enhancement ratio that is included in the common calculations ad hoc as an empirical factor, appears in the MSA naturally as the appropriate effect of free radicals becomes stronger or weaker depending on the degree of oxygenation. The importance of the scientifically predicted biological outcome of radiation with ions and the following calculation of the desired pattern of physical dose distribution along the SOBP for reliable planning of proton therapy cannot be overestimated. We are confident that the MSA-based analysis will give an effective way to relate the physical dose delivered with protons to the probability of cell survival.

%Refs.\cite{pre,MSAColl,Radicals,CellSurSR16}. The main success of this approach regarding this aim is that there is an understanding of how to predict the probability of cell survival based on the dose, LET, and oxygen conditions of the cell surroundings.

The goal of this work is to develop such an algorithm, but in order to reach this goal, we first need to calculate the survival probability at an arbitrary position within the SOBP. After this is achieved, we will calculate the profile of cell survival probability along the SOBP for a uniform profile of physical dose. Then, the profile of physical dose that makes the cell survival probability along the SOBP uniform will be found.

In Refs.~\cite{MSAColl,CellSurSR16} the cell survival probabilities were calculated as function of dose, LET, and oxygen concentration.
%These probabilities are calculated for a given type of cell, at a given LET, oxygenation conditions, and a given dose.
%%%%%%%%%%%%%%%%%%%%%%%%%%%%%%%%%%%%%%%%%%%%%%%%%%%%%%%%%%%%%%%%%%%%%%%%%%%%%%%%%%%%%%%%%
A clustered DNA lesion is defined as the number of DNA lesions, such as DSBs, SSBs, abasic sites, damaged bases, etc., that occur within about two helical turns of a DNA molecule so that, when repair mechanisms are engaged, they treat a cluster of several of these lesions as a single damage site~\cite{Ward2,Iliakis13,Ward1,Goodhead94,Lynn1,Lynn2,Lynn11}.
In ref.~\cite{MSAColl}, a criterion for lethality of damage was suggested and implemented for the calculation of survival curves. This criterion is based on the idea that among different DNA lesions caused by interaction with reactive species the multiply damaged sites with a sufficient complexity may not be repaired (this followed Ref.\cite{precomplex}).
Namely, it was postulated~\cite{MSAColl} that a lesion consisting of a DSB and at least two other simple lesions such as SSB within two DNA twists is lethal. Then in Ref.~\cite{CellSurSR16} this criterion was applied and justified for a number of cell types.

The first step after the justification of the criterion of lethality is to calculate the number of lethal damage sites, produced by an ion (with a given energy and, hence, LET) per length of a segment of its trajectory, $\frac{d N_l}{d x}$ (where $dx$ is a length of this segment)~\cite{MSAColl}. That calculation relies on the transport of reacting species that included a collective flow due to the shock waves around ions' paths predicted by the MSA~\cite{prehydro,MSAColl}. The strength of shock waves and radius of propagation of reactive species depend on the LET. The collective transport of radicals by shock waves has been most recently demonstrated by MD simulations~\cite{PabloOH};  for these simulations, the MBN Explorer with reactive forcefield~\cite{MBNX,MBNXCh} has been employed. In principle, the value of $\frac{d N_l}{d x}$ can be studied experimentally by the analysis of repair foci~\cite{Jakob} and we hope that a more thorough comparison with experimental results will be done, but this is only a side issue for this work.

The value of $\frac{d N_l}{d x}$ is proportional to the number density of chromatin, $n_s$, and cross section of lethal damage $\sigma_l(S_e)$,
\begin{eqnarray}
\frac{d N_l}{d x}=n_s \sigma_l(S_e)~.
\label{n.comp}
\end{eqnarray}
{This cross section is discussed in much detail in Refs.~\cite{MSAColl,CellSurSR16}, where it is derived along with the criterion of lethality. The main idea of its derivation is briefly outlined in the appendix.}
Strictly speaking, besides LET (so-called restricted LET, $S_e$, that excludes nuclear fragmentation effects), $\sigma_l(S_e)$ depends on the concentration of oxygen~\cite{CellSurSR16}. The oxygen concentration influences the effectiveness of reactive species. In this work, we assume that this concentration does not vary along the SOBP, but in principle it may and this can be included in the method suggested below. Then the average yield of lethal DNA lesions in the cell nucleus is given by the product of $\frac{d N_l}{d x}$ in (\ref{n.comp}) and the average length of traverse of all ions passing through a cell nucleus for a given dose~\cite{MSAColl,CellSurSR16},
\begin{equation}
Y_{l}
= \frac{{d}N_{l}}{{d}x} \, \bar{z} \, N_{\rm ion}(d),
\label{eq08}
\end{equation}
where the length of traverse is presented by a product of average length of traverse by a single ion $\bar{z}$ and the number of ions that pass through the cell nucleus $N_{\rm ion}$. The latter depends on dose $d$ as
\begin{equation}
N_{\rm ion} = A_{\rm n} \, d / S_{\rm e} \ ,
\label{eq_031}
\end{equation}
where $A_{\rm n}$ is the cross sectional area of the cell nucleus. Equations~(\ref{eq08}) and (\ref{eq_031}) can be combined:
\begin{eqnarray}
Y_{l}
= \frac{{d}N_{l}}{{d}x} \, \bar{z} \, N_{\rm ion}(d)
= \frac{\pi}{16} \sigma_l(S_e) \, N_{\rm g} \, \frac{d}{S_e} \nonumber \\= \frac{\pi}{16} \sigma_l(S_e) \, N_{\rm g} \frac{N_{\rm ion}}{A_{\rm n}}  = \frac{\pi}{16} \sigma_l(S_e) \, N_{\rm g} F ~,
\label{eq08a}
\end{eqnarray}
where $N_{\rm g}$ is the genome size, i.e., a constant number and $\frac{N_{\rm ion}}{A_{\rm n}}=F$ is the ions' fluence in the beam.

Since the probability of cell inactivation is obtained by subtracting the probability of zero lethal lesions occurrence from unity,
($1-\exp\left[-Y_{l}\right]$) and that of cell survival is given by unity less the probability of cell inactivation, the logarithm of cell survival probability is simply given by Eq.~(\ref{eq08a}) with a negative sign, i.e.,
\begin{equation}
\ln{\Pi_{surv}}=-Y_{l}=- \frac{\pi}{16} \sigma_l(S_e) \, N_{\rm g} F~.
\label{logprob.surv}
\end{equation}
This formula, verified in Ref.~\cite{CellSurSR16}, is remarkable: it suggests that cell survival at a given place depends only on LET, oxygen concentration (both contained in $\sigma_l(S_e)$), and the ions' fluence.

If the Bragg peak is not pristine, Eq.~(\ref{eq_031}) has to be modified, because the ion's fluence is not a number any more, but rather a superposition of fluences of ions with different energies.

Instead there will be a superposition of numbers of ions traversing a cell nucleus at a different energies with different LET. Equation~(\ref{logprob.surv}) will change as well:
\begin{eqnarray}
\ln{\Pi_{surv}}=-\frac{\pi}{16} N_{\rm g} \sum_j \sigma_l(S_{j})
F_j~,
\label{logprob.surv.sobp}
\end{eqnarray}
where index $j$ corresponds to a given fraction of ion's fluence corresponding to a certain energy and LET within a SOBP.

%%%%%%%%%%%%%%%%%%%%%%%%%%%%%%%%%%%%%%%%%%%%%%%%%%%%%%

In order to reproduce the problem with nonuniform survival probability along the SOBP, we need to construct the SOBP, i.e., find such a linear combination, $\sum_j f_j F$ (where $f_j$ are coefficients corresponding to ions with initial kinetic energy $E_{0,j}$), that will produce a uniform dose distribution along the SOBP. In order to do this, the depth-dose curve for protons (in this case) has to be known. There are many more or less sophisticated ways to obtain these curves, e.g., based on Monte Carlo simulations of ions' transport~\cite{Molina,FriedlandSR2017}. However, this work is methodological, and we use a calculation scheme~\cite{epjd,Surdutovich09_PRE,MSAColl}, where the shape of a Bragg peak including energy straggling and charge transfer has derived from a semi-empirical model~\cite{Rudd92}. The result is the dose-depth distribution for a pristine Bragg peak. Then stepping down in protons initial energy by 0.5 MeV, we can calculate what fraction of a beam of this energy is needed in order to achieve a uniform dose distribution. This procedure lasts through the full length of the SOBP.
%%%%%%%%%%%%%%%%%%%%%%%%%%%%%%%%%%%%%%%%%%%%%%%%%%%%%%
%and $S_{\rm e} = |{d}E/{d}x|$ is a part of LET spent on ionization of tissue. This gives us the dose dependence.

The dose at a given depth $x$ is proportional to the net LET:
\begin{eqnarray}
S_e(x) = \sum_j f_j S_j(x) dx=S_0~,
\label{dosex}
\end{eqnarray}
where $S_0$ is a constant. The stopping power at a distal side of the SOBP is equal to the $S(E_{max})$, therefore $S_0=S(E_{max})$. We can start the construction from the distal side:
\begin{eqnarray}
S_e(x_0) = S_0(E_0,x_0)\nonumber \\
S_e(x_1) = S_0(E_0,x_1)+f_1 S_1(E_0-\Delta E, x_1)\nonumber \\
S_e(x_2) = S_0(E_0,x_2)+f_1 S_1(E_0-\Delta E, x_2)\nonumber \\ +f_2 S_2(E_0-2\Delta E, x_2)\nonumber \\
...\nonumber \\
S_e(x_N) = \sum f_j S_j(E_0-N \Delta E, x_N)~,
\label{dosex1}
\end{eqnarray}
where $x_j$ is the coordinate of the Bragg peak corresponding to initial energy $E_{0,j}=E_0-j\Delta E$
For each $j$, the coefficient $f_j$ is determined from equation $j$. The length of the SOBP, $x_p$, is equal to the difference $x_0-x_N$. This condition determines $N$. The example of such a construction is given in Fig.~\ref{fig.bragg}, where the SOBP is constructed for $x_p=12$~mm with $\Delta E=0.5$~MeV (this number is used in current clinical protocols).

\begin{figure}
\begin{centering}
\resizebox{0.99\columnwidth}{!}
{\includegraphics{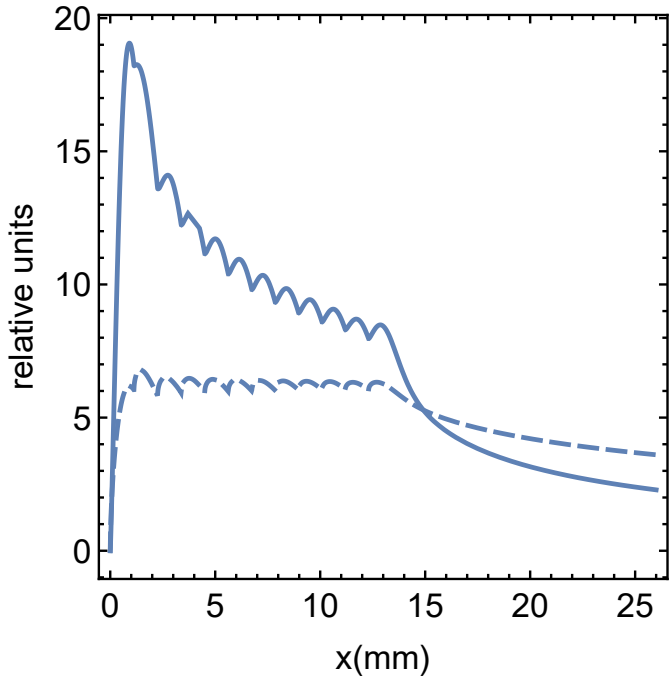}}
\caption{\label{fig.brush1} The solid line shows the profile of dependence of yield of lethal lesions in cells along the SOBP as a function of distance on the distal end of the SOBP. The dashed line shows the profile of the depth-dose curve that produced the above result. The algorithm (\ref{dosex1}) has been used.}
\end{centering}
\end{figure}

\begin{figure}
\begin{centering}
\resizebox{0.99\columnwidth}{!}
{\includegraphics{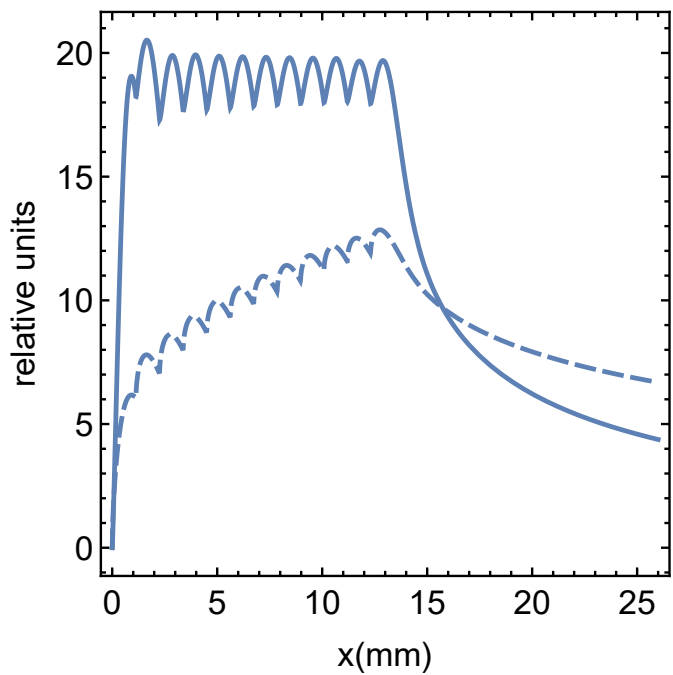}}
\caption{\label{fig.brush2} The solid line shows the profile of dependence of yield of lethal lesions in cells along the SOBP as a function of distance on the distal end of the SOBP. The algorithm (\ref{dosex2}) has been used. The dashed line shows the profile of the depth-dose curve that produced the above result.}
\end{centering}
\end{figure}

Then, the value of $\sigma_l(S_{j})$ can be calculated using the parameters employed in Ref.~\cite{CellSurSR16}. The results for survival probabilities calculated using Eq.~(\ref{logprob.surv.sobp}) are shown in Fig.~\ref{fig.brush1}. The dose profile is shown with a dashed curve in order to better illustrate the problem: a uniform distribution of the physical dose along the SOBP yields a nonuniform distribution of cell survival. A simple explanation of this problem is that (as has been stated above) the radiation damage with ions is not just the product of dose, but the LET and physical consequences of high LET in particular. Mathematically, the effect due to the $\sigma_l(S_{j})$ dependence on LET. Largely, there two effects define $\sigma_l(S_{j})$. One is the number reactive species including secondary electrons. This number is roughly proportional to the LET. If this would be the only effect, then, according to Eq.~(\ref{eq08a}), the average yield of lethal DNA lesions would be proportional to the dose only. If the uniform physical dose is delivered using the corresponding choice of coefficients $f_j$, as the dashed line shows in Fig.~\ref{fig.brush1}, the yield would be uniform (as the solid line shows in Fig.~\ref{fig.brush2}). The other effect is the effective spreading of the reactive species, discussed in relation with the predicted shock waves~\cite{natnuke,MSAColl,Radicals} and demonstrated in Ref.~\cite{PabloOH}. This effect is more complex, but is linear in the first order with respect to LET.
%This is the sense of the dependence shown in Fig.~\ref{fig.brush1}.
This argument explains why this problem has not been solved by the track structure community. Since all their radiation damage estimates are based on the deposited energy and the collective transport by the shock wave is missing, the predicted damage is proportional to the dose while the LET dependence is missing.

Now we can solve an inverse problem: aiming at a given uniform survival probability, find the coefficients $f_j$ for initial ions' fluences distribution. For that, we need to keep $\sum_j f_j \sigma_l(S_{j})$ constant along the SOBP and find the coefficients $f_j$ from this condition.
Similarly to Eq.(\ref{dosex1}) we write:
\begin{eqnarray}
\sigma_l(S_e(x_0)) = \sigma_l(S_0(E_0,x_0))\nonumber \\
\sigma_l(S_e(x_1)) = \sigma_l(S_0(E_0,x_1))+f_1 \sigma_l(S_1(E_0-\Delta E, x_1))\nonumber \\
\sigma_l(S_e(x_2)) = \sigma_l(S_0(E_0,x_2))+f_1 \sigma_l(S_1(E_0-\Delta E, x_2))\nonumber \\ +f_2 \sigma_l(S_2(E_0-2\Delta E, x_2))\nonumber \\
...\nonumber \\
\sigma_l(S_e(x_N)) = \sum f_j \sigma_l(S_j(E_0-N \Delta E, x_N))~,
\label{dosex2}
\end{eqnarray}
Similarly, for each $j$, the coefficient $f_j$ is determined from equation $j$. The results of application of this algorithm are shown in Fig.~\ref{fig.brush2}; again, the dose profile is shown with a dashed curve for guidance, while the solid line shows a profile of the yield of lethal lesions. In addition to this, we want to add, that the oxygen effect can also be added to the calculation of $\sigma_l(S_{j})$ on the local basis. Then it may also affect the choice of $f_j$s. Thus, the problem has been solved.

In conclusion, the problem of varying cell survival along the SOBP has been investigated using the MSA. Its origin has been related to the physical effects such as propagation of reactive species by collective flows induced by ion-induced shock waves that take place at high LET. The solution to this problem is suggested and it may lead to a substantial improvement of treatment planning. The suggested method calls for further investigations, especially experimental. A thorough study of cell survival along the SOBP is strongly desirable. Theoretically, a better depth-dose curves should be used for practical applications, however, as it has been explained, the cause of the effect has nothing to do with the shape of these curves.

The suggested algorithm can be implemented in novel treatment planning and optimisation codes. Differences in cells and conditions can be included in the existing method. Manifestations of other biological phenomena, such as DNA repair beyond linear effect~\cite{CellSurSR16}, bystander effect, etc. can be included empirically until a fundamental understanding is achieved.

\begin{appendix}
\section*{Appendix: Calculation of $\sigma(S_e)$}

As is explained in the main text,
\begin{eqnarray}
\sigma(S_e)=\frac{1}{n_s}\frac{d N_c}{d x}~,
\label{a.sigma}
\end{eqnarray}
is the cross section of inducing a lethal damage in a cell nucleus. Its calculation has been discussed in detail in Refs.~\cite{MSAColl,CellSurSR16} and we are only briefly go over it.

The calculation starts with the determining a number of secondary electrons incident on a target, defined as a site on which a lethal damage can be induced. Lethality is introduced as a criterion of minimal complexity/multipli\-city of lesions within a site that make it irreparable biologically. The number of lesions is proportional to the number of incident electrons, while the latter is proportional to LET. Then, the concentration of reactive species multiplied by their effectiveness is introduced. Both the number and effectiveness depend on the chemical properties of the medium that include the abundance of oxygen (both increase with increasing concentration of oxygen).

The average number for multiply damage sites (per target) containing clustered damage at a distance $r$ from the ion's path is given by
\begin{eqnarray}
{\cal N}_c(r) ={\cal N}_e(r)+{\cal N}_r(r) = \Gamma_e {\cal F}_e(r)+\Gamma_r {\cal F}_r(r)~,
\label{nrhocal}
\end{eqnarray}
where the functions ${\cal N}_e(r)$ and ${\cal N}_r(r)$ define the average number of lesions like SSBs, base damages, abasic states, etc., induced by secondary electrons and other reactive species (free radicals, pre-solvated and solvated electrons, etc.), respectively. ${\cal F}_e(r)$ is the number of secondary electrons incident in the target at a distance $r$ from the ion's path. ${\cal F}_r(r)$ is that for other reactive species. $\Gamma_e$ and $\Gamma_r$ are the probabilities of inducing simple lesions to a DNA molecule by the corresponding species on impact.

The criterion of lethality described above is introduced as follows. The probability of lethal damage, $P_l(r)$, is given by
\begin{eqnarray}
P_l(r)=\lambda \sum_{\nu=3}^\infty\frac{{\cal N}_c^\nu}{\nu !}\exp{\left[-{\cal N}_c\right]}~.
\label{nrob.comp}
\end{eqnarray}
The sum starts from $\nu=3$, which makes the minimum order of lesion complexity at a given site larger or equal to three. The factor $\lambda$ indicates that one of the simple lesions is converted to a DSB. This implies that in the current model the DSBs occur via SSB conversion; in principle, other mechanisms can also be taken into account.

Then the cumulative effect of secondary electrons and reactive species have to be integrated over the volume for a given segment of ion's trajectory. At this point, the radial range to which the reactive species can propagate becomes quite important. This range is determined by the strength of the shock wave whose collective flow is mainly responsible for the propagation of reactive species. According to this, ${\cal N}_r(r)=N_r \theta(R(S_e)-r)$, where $\theta$ is a Heaviside function.
The integral,
\begin{eqnarray}
\frac{d N_l}{d x}= n_s \int_0^\infty P_l(r) 2\pi r d r=n_s \sigma_l(S_e)~,
\label{a.n.comp}
\end{eqnarray}
where $n_s$ is the number density of sites, gives the number of clustered damage sites per unit length of the ion's trajectory, defines $\sigma_l$.
The number density of targets, $n_{\rm s}$, is proportional to the ratio of base pairs
accommodated in the cell nucleus to the nuclear volume, $n_{\rm s} \sim N_{\rm bp}/ V_{\rm n}$~\cite{CellSurSR16}.

Further detail of calculation of ${\cal F}_e(r)$, $\Gamma_e$, etc. can be found in Ref.~\cite{MSAColl}. The latest on the study of $R(S_e)$ is in Ref.~\cite{PabloOH}.
\end{appendix}

\bibliographystyle{epj}
%\bibliography{bibliography1}

\begin{thebibliography}{40}

\bibitem{Kraft07}
U.~Amaldi, G.~Kraft, J. Radiat. Res. \textbf{48}, A27 (2007)

\bibitem{Pedroni05}
M.~Goitein, A.~Lomax, E.~Pedroni, Phys. Today \textbf{55}, 45 (2002)

\bibitem{Hiroshiko}
H.~Tsujii, T.~Kamada, M.~Baba, H.~Tsuji, H.~Kato, S.~Kato, S.~Yamada,
  S.~Yasuda, T.~Yanagi, H.~Kato et~al., New J. Phys. \textbf{10}, 075009 (2008)

\bibitem{SchardtRMP10}
D.~Schardt, T.~Els{\"a}sser, D.~Schulz-Ertner, Rev. Mod. Phys. \textbf{82}, 383
  (2010)

\bibitem{Durante10}
M.~Durante, J.~Loeffler, Nat. Rev. Clin. Oncol. \textbf{7}, 37 (2010)

\bibitem{FokasKraft09}
E.~Fokas, G.~Kraft, H.~An, R.~Engenhart-Cabillic, Biochim. Biophys. Act.
  \textbf{1796}, 216 (2009)

\bibitem{Hall}
E.J. Hall, A.J. Giaccia, \emph{Radiobiology for Radiologist} (Lippincott
  Williams \& Wilkins, Philadelphia, Baltimore, New York, London, 2012)

\bibitem{IBCTbook}
A.~Solov'yov, ed., \emph{Nanoscale Insights into Ion-Beam Cancer Therapy}
  (Springer, 2017)

\bibitem{Alpen}
E.L. Alpen, \emph{Radiation Biophysics} (Academic Press, San Diego, London,
  Boston, New York, Sydney, Tokyo, Toronto, 1998)

\bibitem{Paganetti}
H.~Paganetti et~al., Int. J. Radiation Oncology Biol. Phys. \textbf{53}, 407–
  (2002)

\bibitem{LEM96}
M.~Scholz, G.~Kraft, Adv. Space Res. \textbf{18}, 5– (1996)

\bibitem{Kraemer}
T.~Elsaesser, M.~Kraemer, M.~Scholz, Int. J. Radiat. Oncol. Biol. Phys.
  \textbf{71}, 866 (2008)

\bibitem{Elsaesser}
M.~Scholz, A.~Kellerer, W.~Kraft-Weyrather, G.~Kraft, Radiat. Environ. Biophys.
  \textbf{36}, 59 (1997)

\bibitem{Beuve}
M.~Beuve, Radiat. Res. \textbf{172}, 394– (2009)

\bibitem{MSAColl}
E.~Surdutovich, A.~Solov'yov, Eur. Phys. J. D \textbf{68}, 353 (2014)

\bibitem{LEMcluster}
T.~Elsaesser, M.~Scholz, Radiat. Res. \textbf{167}, 319 (2007)

\bibitem{lem2012}
T.~Friedrich, U.~Scholz, T.~Els{\"a}sser, M.~Durante, M.~Scholz, Int. J.
  Radiat. Biol. \textbf{88}, 103– (2012)

\bibitem{Paganetti16}
T.~Underwood, H.~Paganetti, Int. J. Radiation Oncology Biol. Phys. \textbf{95},
  56– (2016)

\bibitem{pre}
A.~Solov'yov, E.~Surdutovich, E.~Scifoni, I.~Mishustin, W.~Greiner, Phys. Rev.
  \textbf{E79}, 011909 (2009)

\bibitem{CellSurSR16}
A.~Verkhovtsev, E.~Surdutovich, A.~Solov'yov, Sci. Rep. \textbf{6}, 27654
  (2016)

\bibitem{Ward2}
J.~Ward, Radiat. Res. \textbf{142}, 362 (1995)

\bibitem{Iliakis13}
A.~Schipler, G.~Iliakis, Nucl. Acid. Res. \textbf{41}, 7589– (2013)

\bibitem{Ward1}
J.~Ward, Prog. Nucleic Acid. Res. Mol. biol. \textbf{35}, 95 (1988)

\bibitem{Goodhead94}
D.T. Goodhead, Int. J. Radiat. Biol. \textbf{65}, 7 (1994)

\bibitem{Lynn1}
S.~Malyarchuk, R.~Castore, L.~Harrison, DNA Repair \textbf{8}, 1343 (2009)

\bibitem{Lynn2}
S.~Malyarchuk, R.~Castore, L.~Harrison, Nucleic Acids Res. \textbf{36}, 4872
  (2008)

\bibitem{Lynn11}
E.~Sage, L.~Harrison, Mutat. Res. \textbf{711}, 123 (2011)

\bibitem{precomplex}
E.~Surdutovich, D.C. Gallagher, A.V. Solov'yov, Phys. Rev. E \textbf{84},
  051918 (2011)

\bibitem{prehydro}
E.~Surdutovich, A.~Solov'yov, Phys. Rev. E \textbf{82}, 051915 (2010)

\bibitem{PabloOH}
P.~de~Vera et~al., Phys. Rev. Lett. \textbf{submitted} (2017)

\bibitem{MBNX}
I.A. Solov'yov, A.V. Yakubovich, P.V. Nikolaev, I.~Volkovets, A.V. Solov'yov,
  J. Comp. Chem. \textbf{33}, 2412 (2012)

\bibitem{MBNXCh}
G.~Sushko, I.~Solov'yov, A.~Verkhovtsev, S.V. Volkov and A.V.~Solov'yov, Eur. Phys. J.
  D \textbf{70}, 12 (2016)

\bibitem{Jakob}
B.~Jakob, M.~Scholz, G.~Taucher-Scholz, Radiat. Res. \textbf{159}, 676 (2003)

\bibitem{Molina}
I.~Abril, R.~Garcia-Molina, C.~Denton, I.~Kyriakou, D.~Emfietzoglou, Radiat.
  Res. \textbf{175}, 247 (2011)

\bibitem{FriedlandSR2017}
W.~Friedland, E.~Schmitt, P.~Kundrat et~al., Sci. Rep. \textbf{7}, ??? (2017)

\bibitem{epjd}
E.~Surdutovich, O.~Obolensky, E.~Scifoni, I.~Pshenichnov, I.~Mishustin,
  A.~Solov'yov, W.~Greiner, Eur. Phys. J. D \textbf{51}, 63 (2009)

\bibitem{Surdutovich09_PRE}
E.~Scifoni, E.~Surdutovich, A.~Solov'yov, Phys. Rev. E \textbf{81}, 021903
  (2009)

\bibitem{Rudd92}
M.E. Rudd, Y.K. Kim, D.H. Madison, T.~Gay, Rev. Mod. Phys. \textbf{64}, 441
  (1992)

\bibitem{natnuke}
E.~Surdutovich, A.V. Yakubovich, A.V. Solov'yov, Sci. Rep. \textbf{3}, 1289
  (2013)

\bibitem{Radicals}
E.~Surdutovich, A.~Solov'yov, Eur. Phys. J. D \textbf{69}, 193 (2015)

\end{thebibliography}

\end{document}